\documentstyle[prl,aps,amsfonts]{revtex}

\newcommand{\re}{\mathop{\mathrm{Re}}}
\newcommand{\im}{\mathop{\mathrm{Im}}}
\newcommand{\tr}{\mathop{\mathrm{tr}}}
\newcommand{\diag}{\mathop{\mathrm{diag}}}

\begin{document}

\draft

\twocolumn[\hsize\textwidth\columnwidth\hsize\csname@twocolumnfalse%
\endcsname

\title{Systematic analytical approach to correlation functions of resonances in
quantum chaotic scattering}

\author{
Yan V. Fyodorov$\S$ and Boris A. Khoruzhenko$\P$}
\address{
$\S$Fachbereich Physik, Universit\"at-GH Essen,
D-45117 Essen, Germany
        }
\address{
$\P$ School of Mathematical Sciences, Queen Mary \& Westfield
College, \\ University of London, London E1 4NS, U.K.
        }
\date{May 6, 1999}
\maketitle

\begin{abstract}
We solve the problem of resonance statistics in systems with
broken time-reversal invariance by deriving the joint probability
density of all resonances in the framework of a random matrix
approach and calculating explicitly all n-point correlation
functions in the complex plane. As a by-product, we establish the
Ginibre-like statistics of resonances for many open channels. Our
method is a combination of Itzykson-Zuber integration and a
variant of nonlinear $\sigma-$model and can be applied when
the use of orthogonal polynomials is problematic.
\end{abstract}

\pacs{PACS numbers: 05.45.+b} ]
As is well-known, universal statistical properties of bound states
in the regime of quantum chaos can be described in the framework of
the random matrix approach \cite{Bohigas}. The relevant methods adjusting
random matrix description to the case of
resonance scattering in open quantum systems are very well-known
since the pioneering work by the Heidelberg group\cite{VWZ}, see the
review \cite{FSR} for a thorough discussion of recent
developments.

One of the basic concepts in chaotic quantum scattering is the
notion of resonances. Resonances are
 long-lived  intermediate states
to which bound states of a ``closed'' system are converted due to
coupling to continua.
On the formal level the resonances show up as
poles of the $M\times M$ scattering matrix $S_{ab}(E)$.
The dimension of this matrix, $M$, equals the number of
open channels in a given interval of energies. The poles of
$S_{ab}(E)$ occur at complex energies
${\cal E}_k\!=\!E_k\!-\!\case{i}{2}\Gamma_k$,
where $E_k$ is called the position and $\Gamma_k$ the width of
the corresponding resonance. Recent advances in
computational techniques made available
resonance patterns with high accuracy  for realistic
models of atomic and molecular chaotic systems\cite{Blumel},
as well as for quantum billiards and other models
related to chaotic scattering \cite{unpub}.

In the framework of the random matrix approach the S-matrix poles
(resonances) are just the complex eigenvalues of an effective
random matrix Hamiltonian ${\cal H}_{\text{eff}}\!=\!H\!-\!
i\Gamma$. Here $H$ is a random self-adjoint matrix of a large dimension $N$
describing the statistical
properties of the {\it closed} counterpart of the scattering
system under consideration. Depending on presence or absence of the
time-reversal invariance $H$ has to be chosen as a real symmetric or complex
Hermitian one, respectively\cite{Bohigas,Mehta}.
The $N\!\times\! N$ matrix $\Gamma$ serves
the purpose of describing
transitions from the states described
by $H$ to the outer world via $M$ open channels. It is
related to
the $N\!\times\! M$ matrix $W$ of transition amplitudes in the following way:
$\Gamma\!=\!\pi WW^{\dagger}$. Such a form of $\Gamma$ is
actually dictated by
the requirement of the $S-$matrix unitarity and ensures that all
$S-$matrix poles lie in the lower half-plane of complex energies,
as required by causality. It is evident that the rank of
$\Gamma$ is $M$. In practice, the most interesting case is that of few
open channels: $N\gg M\sim 1$. In this case the width $\Gamma_k$ of a
typical resonance
is comparable with the mean {\it separation} $\Delta$ between
neighboring resonances along the real axis and statistical
properties of resonances are expected to be universal\cite{FSR}.

Despite quite substantial efforts\cite{Sok,FSR,SFT,FTS} our
actual knowledge of S-matrix poles statistics for few-channel
scattering is still quite restricted. Among the facts
established analytically beyond perturbation theory
one can mention (i) the density of joint distribution of all resonances
for the system with a single open channel and gaussian-distributed
transition amplitudes $W$ \cite{Sok} and  (ii) the mean density
of S-matrix poles for arbitrary $M\!\ll\! N$ \cite{FSR,SFT}, as well as 
for $M\!\sim\! N$\cite{Haake}.

At the same time,
the most interesting and difficult question of correlations
between resonances in the complex plane
resisted systematic analytical investigations.
As an attempt to get an insight into the problem a nontrivial integral
relation satisfied by the lowest (two-point) correlation
function of complex eigenvalues for $\Gamma\geq 0$
was derived recently in \cite{FTS}. Using that relation it turned out to be
possible to put forward a conjecture on
the analytic structure of the correlation functions for systems with broken
time-reversal invariance.
Unfortunately, the above mentioned relation neither fixed the
lowest correlation function in a unique way nor provided a
direct information on higher correlation functions.

The goal of the present paper is to develop a regular analytical
approach to the statistical properties of resonances for systems with
broken time-reversal invariance. To this end, we first derive the joint
probability density of all resonances for arbitrary number of open channels
$M$. Then we reduce the problem of extracting the $n-$point
correlation functions in the limit $n,M$-fixed, $N\to \infty$ to averaging
a certain product of $2n$ determinants. Finally, the latter is evaluated
with a method combining a mapping to a fermionic
version of a nonlinear $\sigma-$model with the Itzykson-Zuber integration.
As a result, we
prove the validity of the conjecture put forward in \cite{FTS}.

We consider an ensemble of random $N\times N$ complex matrices
${J}={H}+i{\Gamma}$\cite{note}, where ${H} $ is $N\times N$ matrix taken from
a Gaussian Unitary Ensemble (GUE) of {\it Hermitian } matrices
with the probability density ${\cal P}({H})\propto
\exp{(-\case{N}/{2} \tr {H}^2)}$ and $\Gamma$ is a 
{\it fixed} nonnegative one: $\Gamma\ge 0$.
The probability density function in our ensemble  
can be written in the form
\begin{equation}\label{P(J)}
{\cal P}({J})\propto \exp \left[ -\frac{N}{2} \tr\,
\left(\frac{J+J^{\dagger}}{2}\right)^{\!\!2}\right] \delta \left(\Gamma
- \frac{J-J^{\dagger}}{2i}\ \right).
\end{equation}
Here and henceforth we do not specify the multiplicative constants
 when dealing with probability
densities and correlation functions since
they can be always found from the normalization condition.

Eq.\ (\ref{P(J)}) can be used to obtain the density of joint
distribution of eigenvalues by integrating ${\cal P}(J)$ over the
degrees of freedom that are complementary to the eigenvalues of
$J$. This can be done following Dyson's method (see \cite{Mehta,FKS}).
Neglecting matrices with repeated
eigenvalues, one transforms $J$ to triangular form:
${J}={U}({Z}+{R}){U}^{\dagger}$, where ${U}$ is a unitary matrix,
${R}$ is strictly upper-triangular and ${Z}=\diag\{ z_1, \ldots
z_N \}$ is the diagonal matrix of complex eigenvalues. 
The Jacobian of the transformation from $J$ to $(Z, U, R)$ 
is $|\Delta(\!{Z})|^2$ where
$\Delta(\!{Z})=\prod_{1\le j<k\le N}(z_j-z_k)$ is the Vandermonde
determinant. To perform the integration over ${R}$ it is
technically convenient to use a Fourier-integral representation
for the $\delta-$function in Eq.\ (\ref{P(J)}). This reduces the
corresponding integral to a Gaussian one and after algebraic
manipulations the resulting expression is
\begin{equation}\label{P(Z)}
{P}_{N, \Gamma} ({Z})\propto e^{-\frac{N}{2}  \re \tr {Z}^2
-\frac{N}{2}\tr {\Gamma}^2 }|\Delta (\!{Z})|^2{Q}(\im {Z} ),
\end{equation}
where $ {Q}(\im {Z} ) $ is the remaining integral
\begin{equation}\label{y:1}
{Q}(\im {Z} )=\int\! {[dU]} \prod_{l=1}^N\delta
\left(\im{z}_l-({U}^{\dagger}{\Gamma}{U})_{ll}\right),
\end{equation}
over the unitary group $U(N)$, $[dU]$ being the Haar measure. Due to
the specific structure of the matrices $J$ their eigenvalues lie
in the upper part of the complex plane and in all formulae below
 $\im z_j\ge 0$ for all $j$\cite{note}.

To proceed further we need to integrate over $U$. Again it is
convenient to use the Fourier-integral representation for the
$\delta-$functions in Eq.\ (\ref{y:1}):
\begin{equation}\label{y:2}
{Q}(\im {Z} ) = \int\!\!\!\frac{dK}{(2\pi)^N}\
 e^{i\, \im \tr{K}{Z} }
\!\!\int\!{[dU]}\ e^{-i\, \tr {K} {U}^{\dagger}{\Gamma} {U} },
\end{equation}
where the first integration is over all real diagonal matrices
${K}$ of dimension $N$, $dK$ being $dk_1 \ldots dk_N$.

When the eigenvalues of ${\Gamma}$ are all \emph{distinct} the
integration over $U$ can be performed using the famous
Itzykson-Zuber-Harish-Chandra (IZHC) formula\cite{IZHC}. We,
however, are mostly interested in the case when ${\Gamma}$ has a
small rank $M\ll N$, i.e.\  it has only $M$ nonzero eigenvalues
which we denote by $\gamma_1, \ldots , \gamma_M$. This limit of
highly degenerate eigenvalues is difficult to perform in the
original IZHC formula. Nevertheless
the difficulty can be circumvented and the result is as follows:
\begin{eqnarray} \label{start}
{Q}(\im Z)&=&\frac{\det^{M-N} \bbox{\gamma}}
{\Delta(\bbox{\gamma})}\int_{R^M}\!\!d\Lambda\ \det [
e^{-i\gamma_l\lambda_m}]_{l,m=1}^M
\\ & & \times  \Delta(\Lambda ) \prod\limits_{j=1}^N
\sum_{m=1}^M\frac{e^{i\lambda_m \im z_j}} {\prod_{s\ne
m}(\lambda_m-\lambda_s)}\nonumber
\end{eqnarray}
where $\Lambda =\diag (\lambda_1,\ldots,\lambda_M)$ and
$\bbox{\gamma} =\diag (\gamma_1,\ldots,\gamma_M)$.

Eqs.\  (\ref{P(Z)}) and (\ref{start}) give an
explicit representation for the joint probability density of $N$
resonances $z_i$ in the complex plane. As such, they constitute one
of the main results of the present paper and provide the basis
for calculating the $n$-eigenvalue correlation functions
\begin{equation}\label{R_n}
R_n({\sf z})=\frac{N!}{(N-n)!}\! \int\! {\sf dw}\ {P}_{N, \Gamma}
({\sf z},{\sf w}),
\end{equation}
where, for the sake of brevity, we decompose $Z=\diag({\sf z},{\sf w})$ with
${\sf z}\!=\! (z_1, \ldots, z_n)$
and ${\sf w }\!=\! (w_{1}, \ldots, w_{N-n})$, ${\sf
dw}\!=\!\prod_{j=1}^{N-n} d\re w_j\ d\im w_j$, identifying
$w_k\equiv z_{n+k}$.

In what follows we will calculate $R_n({\sf z})$ for arbitrary
fixed $n$ and $M$ in the limit $N\to \infty$. On the first stage
we will replace the integration over ${\sf w}$ in (\ref{R_n}) by
averaging over the ensemble of non-Hermitian random matrices
$J_{N-n}(\bbox{\gamma})=H_{N-n}+i\Gamma$, with $H_{N-n}$ being a
GUE matrix of the reduced size $(N-n)\times (N-n)$. This step
involves cumbersome algebraic manipulations and will be presented
in full details elsewhere. Here we outline the ideas on the
simplest, still nontrivial example of one-channel systems ($M=1$ ). In this
case $\Gamma $ has only one nonzero eigenvalue $\gamma $ 
and the $\lambda-$integration in Eq.\ (\ref{start}) can
be explicitly performed yielding
\begin{equation}\label{M1}
{P}_{N, \gamma} ({Z})\! \propto \frac{|\Delta
(\!{Z})|^2}{\gamma^{N-1}}\ e^{-\frac{N}{2}[\re \tr
{Z}^2+\gamma^2]}\ \delta\Big(\!\gamma -\!\sum_{j=1}^N\im
z_j\!\Big).
\end{equation}
Introducing the notation $\tilde \gamma = \gamma-\sum_{j=1}^n\im
z_j$, we write the $\delta$-function in Eq.\ (\ref{M1}) as
$\delta\left(\tilde{\gamma}-\sum_{l=1}^{N-n}\im w_l\right)$. Now
we replace $N$ in the exponent of Eq.\ (\ref{M1}) by $N-n$ (this
act is justified by the limit $N\to \infty$). With the resulting
relation in hand one readily obtains that
\begin{eqnarray}\label{F}
R_n({\sf z}) &\propto& \frac{C_{\tilde \gamma }({\sf
z})}{\gamma^n} | \Delta({\sf z})|^2 e^{-\frac{N-n}{2}\sum_{j=1}^n
\re\,z_j^2}
\\ \nonumber & & \times \left[
\frac{\tilde{\gamma}}{\gamma}\right]^{N-n-1}\!\! \!\!\!
e^{-\frac{N-n}{2}\left(\gamma^2-\tilde{\gamma}^2\right)},
\end{eqnarray}
where
\begin{eqnarray}\nonumber
C_{\tilde \gamma }({\sf z}) &=& \int\!\!{\sf dw}\
 {P}_{N-n, \tilde \gamma
}({\sf w}) \ \prod_{l=1}^{N-n}\prod_{j=1}^n|z_j-w_l|^2
\\ &= & \left\langle\prod_{j=1}^n
\Big| \det
\left[z_j-{J}_{N-n}(\tilde{\bbox{\gamma}})\right]\Big|^2
\right\rangle_{GUE} \label{C}
\end{eqnarray}

In the limit when $N\to\infty$ and $M$ is finite (in particular, for the present case $M=1)$, the imaginary 
part of almost all 
eigenvalues of $J$ is of the order $\frac{1}{N}
\ll \gamma_m$\cite{note1}, hence so is $\tilde \gamma -\gamma $. 
Therefore one can reinstate $\gamma $ in place of $\tilde \gamma $ 
in the determinants in Eq.\ (\ref{C}).  On the other hand, 
rescaling the imaginary parts $y_j=N\im
z_j$, one finds that 
\[
\left[ \frac{\tilde{\gamma}}{\gamma}\right]^{N-n-1}\!\! \!\!\!
e^{-\frac{N-n}{2}\left(\gamma^2-\tilde{\gamma}^2\right)}=
e^{-2\sum_{j=1}^n y_j g}
\]
in the limit $n\ll N\to \infty$, with
$g=\frac{1}{2}(\gamma+\gamma^{-1})$.

Essentially similar manipulations can be performed for arbitrary
fixed number of open channels $M$. In the limit $n,M\ll N\to
\infty$ we arrive at the following general representation of the
correlation functions:
\begin{eqnarray}\label{interm}
R_n({\sf z})&\propto&  \frac{{C}_{\bbox{\gamma}} ({\sf z})}{\det^n
\bbox{\gamma}}\ |\Delta ({\sf z})|^2\
e^{-\frac{N-n}{2}\sum_{j=1}^n \re\, z_j^2}
\\ \nonumber & & \times
 \prod_{j=1}^n
 \sum_{m=1}^M\frac{e^{-2y_jg_m}}
{\prod_{s\ne m}(g_m-g_s)}
\end{eqnarray}
where $g_m=\frac{1}{2}(\gamma_m+\gamma_m^{-1})$.

Thus, the problem amounts to evaluation of the correlation
function of the determinants in Eq.(\ref{C})\cite{AS}. To proceed, we first
write each of the determinants as a Gaussian integral over a set
of Grassmann variables. When this is done, the GUE average becomes
trivial and yields terms quartic with respect to the
Grassmannians. These  terms can be further traded for an auxilliary
integration over a Hermitian matrix ${S}$ of the size $2n\times
2n$ (the so-called Hubbard-Stratonovich transformation).
Then the integration over the Grassmann fields is trivially
performed and yields again a determinant. As the result, we arrive
at the following expression:
\begin{eqnarray}\label{repr}
{C}_{\bbox{\gamma}}({\sf z})&\propto &\int [ d{S} ] \
 e^{
     -(N-n)\tr
              [
\frac{1}{2}{S}^2 - \ln ({\Bbb Z}_{2n}-i{S})
               ]
    }
\\ \nonumber  & & \times
\prod_{m=1}^M \det \left[{\tt {1\hspace{-0.95ex} 1}}_{2n}+
i\gamma_m {\Bbb L}_{\, 2n} ( {\Bbb Z}_{2n}- i{S})^{-1} \right]
\end{eqnarray}
where we have introduced the diagonal matrices ${\Bbb Z}_{2n} = \diag
({\sf z}, {\sf z}^{\dagger})$ and ${\Bbb L}_{\, 2n}= \diag({\tt
{1\hspace{-0.95ex} 1}}_n,-{\tt {1\hspace{-0.95ex} 1}}_n)$.

Let us now recall that nontrivial eigenvalue correlations are
expected to occur\cite{FKS} on the scale when the eigenvalues are separated
by distances comparable with the mean eigenvalue separation for
GUE matrices $H$, the latter being of the order $(N-n)^{-1}$ with our
choice of ${\cal P}(H) $. Accordingly, it is convenient to separate
the "center of mass" coordinate $x=\frac{1}{n} \sum_{j=1}^n\re
z_j$ so that $z_j = x+\frac{\tilde {z}_j}{N-n}$,
where both the real and imaginary parts of $\tilde{z}_j$ are of the
order of 1 in the limit when $N\to \infty $ and $M$ is
fixed. In this limit $R_n({\sf z})$ is effectively a function of
$\tilde{\sf z}$ ($x$ is fixed) which we are going to calculate.

To evaluate the integral in (\ref{repr}) let us first  diagonalize
$S$: $S={U}_{2n}{\Sigma } U_{2n}^{-1}$, where ${U}_{2n}\in U(2n)$
and ${\Sigma }=\diag (\sigma_1,...,\sigma_{2n})$. Then, keeping
only the leading terms in the limit $N\to \infty$, we obtain:
\begin{equation}\label{CC}
{C}_{\bbox{\gamma}}({\sf z})\!=\!\int\!\! d\Sigma\,
\Delta^2(\Sigma ) e^{-(N-n)\!\!\sum\limits_{k=1}^{2n}[\frac{
\sigma_k^2}{2}- \ln{(x-i\sigma_k)}]}\langle C( {S}
)\rangle_{U(2n)}
\end{equation}
where
\begin{eqnarray*}
\langle C( {S} )\rangle_{U(2n)}&=& \int [d U_{2n}]\  e^{-\tr
[ \tilde{{\Bbb Z}}_{2n}  (x{\tt {1\hspace{-0.95ex} 1}}_{\,2n} - i {S}
)^{-1}]}\\  & & \times \prod_{m=1}^M \det \big[{\tt
{1\hspace{-0.95ex} 1}}_{2n}+ i\gamma_m{\Bbb L}_{2n} (x{\tt
{1\hspace{-0.95ex} 1}}_{2n} - i {S} )^{-1}\big].
\end{eqnarray*}

The form of the integrand in (\ref{CC}) suggests exploiting the
saddle-point method in the integral over $\sigma_k$, $k=1,\ldots,
2n$.  Altogether there are $2^{2n}$ saddle-points:
$\sigma_k^{(s)}=-\frac{i}{2}(x+i\epsilon_k\sqrt{4-x^2})$, where
$\epsilon_k=\pm 1$. The leading order contribution comes from
integration around  those saddle-points where exactly $n$
parameters $\epsilon_k$ equal 1 (the rest being equal -1). All
other choices can be neglected as they lead to lower order terms.
This is because of the presence of the Vandermonde determinant in
the integrand. At the same time, all relevant saddle-points
produce the same contribution and we obtain that:
\begin{equation}\label{CCC}
{C}_{\bbox{\gamma}}({\sf z})\propto
e^{\frac{N-n}{2}\sum_{j=1}^n\re z_j^2}
{C}^{s}_{\bbox{\gamma}}(\tilde{{\sf z}})
\end{equation}
where $\tilde{\sf z}=(\tilde{z}_1, \ldots, \tilde{z}_n)$,
$\tilde{z}_j=N(z_j-x)$, $j=1,\ldots,n$,  and
\begin{eqnarray}\label{c4}
{C}^{s}_{\bbox{\gamma}}(\tilde{{\sf z}})&=& \int [{\rm d} {
\Bbb Q}_{2n}]\  e^{-i\pi\nu(x)\tr\,
\tilde{{\Bbb Z}}_{2n}\!{\Bbb Q}_{2n} }
\\ \nonumber & &
\times \prod_{c=1}^M\det \left[{\tt {1\hspace{-0.95ex} 1}}_{2n}+
\frac{i\gamma_c x}{2}{\Bbb L}_{2n}+ \pi\nu(x)\gamma_c{\Bbb L
}_{2n}{\Bbb Q}_{2n}\right]
\end{eqnarray}
In (\ref{c4}) ${\Bbb Q}_{2n}=U_{2n}^{-1}{\Bbb L}_{\, 2n} U_{2n}$, the
integration is over the coset space $U(2n)/U(n)\!\!\otimes\!\! U(n)$,  and
the symbol $\nu(x)$ stands for the semicircular density of real
eigenvalues of the matrices ${H}$,
$\nu(x)=\frac{1}{2\pi}\sqrt{4-x^2}$.

Thus, the problem reduces to evaluation of an integral over a
coset space. This type of integrals is known in the literature under
the name of zero-dimensional nonlinear $\sigma-$models and our consideration
enjoy many useful insights, see e.g.\cite{zirn}.
In particular, the following polar
parametrization proves to be the most effective:
\[
{\Bbb U}_{2n}\!=\!
  \left(\!\!
      \begin{array}{cc}
          U_A&\\&\!\!U_R
      \end{array}
  \!\!\right)\!\!
  \left(\!\!
      \begin{array}{cc}
        \cos{\hat{\psi}}&e^{i\hat{\phi}}\sin{\hat{\psi}}\\
        e^{-i\hat{\phi}}\sin{\hat{\psi}}&-\cos{\hat{\psi}}
      \end{array}
  \!\!\right)\!\!
  \left(\!\!
      \begin{array}{cc}
         U^{-1}_A&\\&\!\!U^{-1}_R
      \end{array}
  \!\!\right)
\]
where $U_{A,R}\in U(n),\,\, \hat{\psi}=\diag (\psi_1,...,\psi_n)$,
$\hat{\phi}=\diag (\phi_1,...,\phi_n)$ and $0<\phi_k,\psi_k<2\pi$.
The corresponding measure $[{\rm d} {\Bbb Q}_{2n}]$ is
proportional to
\[
[d U_A][dU_R]\prod\limits_{j=1}^n d\phi_j d\psi_j\
\sin{2\psi_j}\!\!\!\prod\limits_{1\le l<k\le n
}|\cos{2\psi_l}-\cos{2\psi_k}|^2.
\]

The use of such a  parametrization makes the integration especially
simple because of the determinant factor being independent of the
unitary matrices $U_{A,R}$. As a result, these matrices appear
only in the exponential factor and the corresponding integrals can
be evaluated according to the IZHC formula\cite{IZHC}. Passing to
the variables $\lambda_k=\cos{2\psi_k}$ we find:
\begin{eqnarray*}\nonumber
{C}^{s}_{\bbox{\gamma}}(\tilde{{\sf z}}) &\propto& \frac{\det^n
\bbox{\gamma} }{|\Delta({\tilde {\sf z}} )|^2}
\int\limits_{-1}^{1}\ \prod_{j=1}^n d\lambda_j
\prod_{j=1}^n G_M(\lambda_j)
\\ \nonumber & & \times
\det \left[
     e^{+i\pi\nu(x)\tilde{z}_j\lambda_k}
     \right]
\det \left[
     e^{-i\pi\nu(x) \tilde{z}_j^*\lambda_k}
     \right]
\\ \nonumber  &\propto& \frac{n!\det^n \bbox{\gamma}}{|\Delta({\tilde {\sf z}}
)|^2}\det \!\! \left[\int\limits_{-1}^{1} \! d\lambda G_M(\lambda)
e^{i\pi\nu(x)\lambda(\tilde{z}_j-\tilde{z}_k^*)}\right]
\end{eqnarray*}
where
\[
G_M(\lambda)=\prod_{m=1}^M [g_m+\pi\nu(x)\lambda]
\]
Combining this with Eqs.\ (\ref{interm}) and (\ref{CCC}) and restoring
the normalization we
finally see that the correlation functions, in the limit $N\to\infty$,
 have the following
simple structure:
\[
\frac{1}{N^{2n}}
R_n\left(x+\frac{\tilde{z}_1}{N}, \ldots ,x+\frac{\tilde{z}_n}{N} \right)=
\det \left[ K (\tilde{z}_j,\tilde{z}_k^*)\right]_{j,k=1}^n,
\]
where the kernel $ K(\tilde{z}_j,\tilde{z}_k^*)$
is given by
\begin{eqnarray}
K(\tilde{z}_1,\tilde{z}_2^*) &=&
F^{1/2}(\tilde{z}_1)F^{1/2}(\tilde{z}_2^*)\\ \nonumber
& & \times \int_{-1}^1  d\lambda
e^{-i\pi\nu(x)\lambda (\tilde{z}_1-\tilde{z}_2^*)}
G_M(\lambda)
\end{eqnarray}
with $F(\tilde{z})= \sum_{m=1}^M\frac{e^{-2|\im \tilde{z}|g_m}} {\prod_{s\ne
m}(g_m-g_s)}$. This is equivalent to the form
conjectured in \cite{FTS}.

One of the important physical limits of the scattering system is
the case of many equivalent open channels: $g_m=g$ for all $m$ and $M\gg g$.
Resonances in that case form a dense cloud in the complex plane
characterized by a  mean density $\rho(z)$ inside the cloud. This
fact and expression for $\rho(z)$ were found in \cite{Haake,FSR}.
Using our formulas derived above we are able to show that the statistics
of the resonances in that case is determined by a {\it Ginibre-like}
kernel:
\begin{equation}\label{Gin}
|K(z_1,z_2)|=\rho(z)\exp{-\frac{1}{2}\pi\rho(z)|z_1-z_2|^2}
\end{equation}
with $z=(z_1+z_2)/2$, generalizing  
a classical result by Ginibre \cite{Gin} to the case of a nonuniform
(i.e. position-dependent) mean density of complex eigenvalues $\rho(z)$.
 As such, it has a good chance to be  universally valid for strongly
non-Hermitian random matrices.  

In conclusion, we considered a non-Hermitian random matrix model of chaotic
quantum scattering. We started with deriving the joint
probability density of all complex eigenvalues
describing S-matrix poles ( resonances) in chaotic systems with
broken time-reversal invariance. Then we found a way to extract all
correlation functions of the resonances in complex plane.
As a by-product, we established the Ginibre-like statistics of resonances for 
many open channels.

Both authors are very
much obliged to O.Zaboronsky for inviting them to
 the Workshop: "Random Matrices and
Integrable Systems" at Mathematical Institute, University of Warwick
discussions at which were instrumental in developing the ideas underlying
the present paper. Y.V.F. acknowledges with thanks
the financial support by SFB-237 and  grant INTAS 97-1342.


\begin{references}
\bibitem{note} We changed the sign in front of $\Gamma$ keeping the condition
$\Gamma\geq 0$ for the sake of notational convenience.
\end{references}
\end{document}